\documentclass[12pt]{article}

\input epsf
\def\beq{\begin{eqnarray}}
\def\eeq{\end{eqnarray}}
\def\ba{\begin{eqnarray}}
\def\ea{\end{eqnarray}}

\def\({\left(}
\def\){\right)}
\def\mn{_{\mu \nu}}

\def\K{{\cal K}}

\def\mpl{M_{\rm pl}}

\def\stu{St\"uckelberg}

\def\lsim{\mathrel{\rlap{\lower3pt\hbox{\hskip0pt$\sim$}}
     \raise1pt\hbox{$<$}}}         
\def\gsim{\mathrel{\rlap{\lower4pt\hbox{\hskip1pt$\sim$}}
     \raise1pt\hbox{$>$}}}         

\newcommand{\comment}[1]{}
\usepackage{amsmath}
\usepackage{amsfonts}
\usepackage{verbatim}
\usepackage{graphicx}
\usepackage{subfigure}
\usepackage{amssymb}
\usepackage[T1]{fontenc}
\usepackage{calligra}

\renewcommand{\comment}[1]{}

\begin{document}

\begin{titlepage}

\thispagestyle{empty}

\begin{flushright}
{NYU-TH-11/08/37}
\end{flushright}
\vskip 0.9cm

\centerline{\Large \bf On Black Holes in Massive Gravity}
\vskip 0.2cm
\centerline{\Large \bf }

\vskip 0.7cm
\centerline{\large L.  Berezhiani$^a$, G. Chkareuli$^a$, C. de Rham$^{bc}$,
G.  Gabadadze$^{ad}$ and A.J. Tolley$^c$}
\vskip 0.3cm

\centerline{\em $^a$Center for Cosmology and Particle Physics,
Department of Physics,}
\centerline{\em New York University, New York,
NY, 10003}

\centerline{\em $^b$D\'epartment de Physique  Th\'eorique and Center for
Astroparticle Physics,  }
\centerline{\em Universit\'e de  Gen\`eve, 24 Quai E. Ansermet, CH-1211  Gen\`eve}

\centerline{\em $^c$Department of Physics, Case Western Reserve University,
Euclid Ave,
Cleveland, OH, 44106}

\centerline{\em $^d$Department of Physics, Columbia University, New York, NY, 10027}

\vskip 1.9cm

\begin{abstract}

In massive gravity the so-far-found black hole solutions on  
Minkowski space happen to convert horizons into a certain type of singularities.
Here we explore whether these singularities can be avoided
if space-time is not asymptotically Minkowskian.
We find an exact analytic black hole (BH) solution  which evades  the above problem
by a transition at large scales to self-induced de Sitter (dS) space-time, with
the curvature scale set by the graviton mass. This solution is similar to the ones
discovered by Koyama, Niz and Tasinato, and  by Nieuwenhuizen, but
differs in detail.  The solution demonstrates that in massive GR, in the
Schwarzschild coordinate system,   a BH   metric has to be accompanied
by the St\"uckelberg fields with nontrivial backgrounds to  prevent
the horizons to convert into the singularities. We also find an
analogous solution for a Reissner-Nordstr\"om  BH  on dS space.
A limitation of our approach, is that  we
find the solutions only for specific values of the two
free parameters of the theory, for which both the vector and scalar
fluctuations loose their kinetic terms, however, we hope our solutions
represent a broader class with better behaved perturbations.

\end{abstract}

\vspace{3cm}

\end{titlepage}

\newpage

\section{Introduction and Summary}

According to the representation theory of the Poincar\'e group in 4D, a massive spin-2 state has to have
five  degrees of freedom; these  can be thought of as the helicity-$0$, $\pm 1$, $\pm 2$ states.
A good Lagrangian for the massive spin-2 has to be able to describe these states.
The Fierz-Pauli mass term \cite{FP} is the only ghost- and tachyon-free term at the quadratic
order that describes the above 5 states \cite{Nieu}. However, in the zero mass
limit it does not recover the linearized Einstein's gravity, since  the helicity-$0$ mode
couples to the trace of the matter stress-tensor with strength equal to that of the helicity-2;
this is called the vDVZ discontinuity \cite{vDVZ}.  If true, it would  rule out massive
gravity on the grounds
of solar system observations. However,  Vainshtein \cite{Arkady} argued that the troublesome
longitudinal mode is suppressed  at measurable  distances by nonlinear effects, making the nonlinear theory
compatible with current empirical data \cite{dvali}. On the other hand,  in a broad class of models,
the same nonlinear terms that are responsible  for the above-mentioned suppression, give rise to an
instability known as the Boulware-Deser (BD) ghost \cite{BD}. This ghost appears as a 6th degree of
freedom in the theory, and even though it    is infinitely heavy on the Minkowski background, it becomes
sufficiently light on locally nontrivial backgrounds, thus invalidating the theory
\cite{nima,Creminelli, DeffayetRombouts,Andrei}.

More recently, however, using  the effective field theory  formalism of
\cite{nima},   it has been  shown  in Ref. \cite{general} that there exists
a two parameter family of nonlinear generalization of the linear
Fierz-Pauli theory, that is free of the BD ghost  order-by order and
to all orders, at least in the decoupling limit.

Most importantly, it was shown in Ref. \cite{resum} that the absence of the
BD ghost in the decoupling limit is
such a powerful requirement that it leads to the resummation of the entire
infinite series of the terms in the effective Lagrangian. As a  result, 
a  candidate theory of massive General Relativity free of BD ghost,
was proposed \cite{resum}.

Using the Hamiltonian analysis  in the unitary gauge 
it was shown that for a certain choice of the free parameters of the
theory,  and in the 4th  order in nonlinearities,  the Hamiltonian constraint that
forbids  the BD ghost is maintained in the theory of \cite{resum}. Note that the quartic order  
is special,  since the lapse necessarily enters nonlinearly in all massive theories 
precisely in this order \cite {Creminelli}, and it may appear that the hamiltonian 
constraint should necessarily be lost then.  In spite of this,  the constraint is maintained in a subtle 
way for special theories, as was shown for a toy model  in \cite {general},  
and shown in the 4th  order for massive GR in \cite{resum}.

The existence of the Hamiltonian constraint to all orders in the unitary gauge,
and  for generic values of the parameters,  was shown in Ref.  \cite{rachel1},
using the method of dealing with the  lapse and shift proposed in
\cite {general,resum}.  Moreover,  Ref.  \cite{rachel1} has
also argued for a secondary constraint, that follows from the conservation of
the Hamiltonian constraint
\footnote{The argument of the existence of the secondary  constraint was challenged 
in \cite {Lham}.}.  Very recently,  the existence of the secondary constraint 
was explicitly confirmed  in Ref.  \cite {Rachel_confirm}.

The absence of the BD ghost among the local fluctuations of 
the theory of \cite {resum} in a generic gauge has been shown  using  the {\stu}  
decomposition \cite {Stu}, as well as the helicity decompositions \cite {Helic} 
to quartic orders in nonlinearities (in the latter two 
references,  previous misconceptions in the literature claiming the presence of the BD 
ghost were also clarified). Motivated by the above developments, in the present  work  
we will proceed  to study certain subtle properties black holes (BH) in the 
theory of \cite {resum}.

In the unitary gauge Lagrangian of  the theory   the object $h\mn \equiv g\mn- \eta\mn$,
is the gravitational analog of the Proca field of  massive electrodynamics, describing all
the five modes of the graviton. The diffeomorphism invariance can be restored by
introducing the four scalar fields
$\phi^a$  (the {\stu} fields) \cite{Siegel,nima,sergey}, and replacing the Minkowski metric by
the covariant tensor $\partial_\mu \phi^a \partial_\nu \phi^b \eta_{ab}$
\beq
g_{\mu\nu}=\partial_\mu \phi^a \partial_{\nu} \phi^b \eta_{ab}+H_{\mu\nu},
\label{H1}
\eeq
where $H\mn$ denotes the covariantized metric perturbation, and $\eta_{ab}={\rm diag} (-1,1,1,1)$.
The existence of the 4 {\stu} scalars $\phi^a$ in this theory leads to the existence of new
invariants in addition to the ones  usually encountered in GR (Ricci scalar, Ricci tensor square,
Riemann tensor square,  etc);  one new basic invariant is
$I^{ab}=g^{\mu\nu}\partial_\mu \phi^a \partial_\nu \phi^b$. Note that the unitary
gauge is set by the condition $\phi^a=x^\mu \delta_\mu^a$.
In this gauge, $I^{ab}=g^{\mu\nu}\delta_\mu^a \delta_\nu^b$. Hence, any inverse
metric that has divergence (even those which are innocuous in GR) would  exhibit
a singularity in the invariant $I^{ab}$. Is this singularity of any significance?
The singularity in the above invariant does not necessarily affect the geodesic motion
of external observers -- the geodesic equation  is identical to that of GR, and due to its
covariance,  one could remove from it what would have been a coordinate
singularity in $g^{\mu\nu}$ in GR.  However, one would expect the singularities in
$I^{ab}=g^{\mu\nu}\delta_\mu^a \delta_\nu^b$ to be a problem for fluctuations around
classical solutions exhibiting it. Since   $g^{\mu\nu}$
could change signs on either side of the singularity, this could lead to emergence
of ghosts and/or tachyons in the fluctuations around a given classical solution.
In what follows, we  will take a conservative point of view and  will
only accept solutions that  have non-singular  $I^{ab}$.  These arguments, in a somewhat
different form,  have already been emphasized recently by Deffayet and Jacobson \cite {CedricJ}.

The above  arguments give  rise to the following seeming puzzle. On the one hand, according to
the Vainshtein mechanism \cite {Arkady}, spherically symmetric solutions of massive gravity
should approximate those of GR better and better,  as we increase the  mass of the source and
come closer to it. This would imply that  the metric of a BH near its horizon should
very much be similar to that of GR. On the other hand, the conventional Schwarzschild metric -- if it
were the solution of massive gravity in unitary gauge -- would be singular according to
the arguments above.

We reiterate this central point in more general terms: In order for a metric to qualify as a valid
description of a BH  configuration, the physical singularities must be absent at the horizon.
Then, in the unitary gauge of massive gravity the  Schwarzschild-like metric
\beq
ds^2=-(1-f)dt^2+\frac{dr^2}{1-f}+r^2d\Omega^2, \quad \text{with e.g.}\quad f=r_g/r,
\label{sh}
\eeq
cannot be a  legitimate BH  solution of the theory.
The same applies to the metric  of de Sitter (dS) space in the static coordinates for which $f=m^2r^2$.

Recently, interesting BH solutions of massive gravity have been found in
Refs. \cite{Koyama,theo,GruzinovMirbabayi} (for other interesting  solutions,
which will not be discussed here, see, \cite{dato}- \cite{LuigiII}).
Following Koyama, Niz and Tasinato (KNT) \cite{Koyama},  one can  start in  the unitary
gauge,   and consider a most general stationary spherically symmetric metric.
Then, using the method developed by KNT,  very interesting full non-linear solutions for stars and black holes
with Minkowskian asymptotics  were found by Gruzinov and Mirbabayi in
\cite{GruzinovMirbabayi}. These solutions  do exhibit the Vainshtein mechanism, and therefore
are potentially  viable classical  solutions  for stars  and other compact objects
in massive gravity (although their stability still remains to be studied).  Nevertheless,
it is not clear, as emphasized in  \cite {GruzinovMirbabayi},  whether these are appropriate
solutions for BHs. Even in  the best case solution, when all the GR invariants are finite,
the invariant $g^{\mu\nu}\partial_\mu \phi^a \partial_\nu \phi^b \eta_{ab}$ diverges,
\cite{GruzinovMirbabayi}. As  noted above, this  divergence does not affect the
geodesic motion of any external observer, however,  we  expect it to be a problem for
fluctuations.

Could  there be any  solution that avoids the above issue? The answer is positive, and the
resolution is in the identification of the unitary gauge to the
coordinate system in which the black hole has no  horizon  (the Kruskal-Szekeres, Eddington-Finkelstein,
or Gullstrand-Painlev\'e systems come to our mind).  The most convenient one  for our purposes
is the Gullstrand-Painlev\'e  (GP) system,  in which the metric has the following form
\beq
ds^2=-dt^2+(d r\pm \sqrt{f}dt)^2+r^2d\Omega^2,
\label{pa}
\eeq
and is free of horizon singularities. It corresponds to the frame of an  in-falling observer
and covers half the whole space  (for either choice of sign).

Furthermore, if one has the metric \eqref{pa} as a solution in unitary gauge, then the coordinate
transformation to the metric \eqref{sh} will lead us to a background with $\phi^a\neq x^a$. This
means that if the configuration is described by the metric \eqref{sh}, the presence of a
halo  of helicity $\pm 1$ and/or $0$  fields  around the BH
is unavoidable. We will show in the present work that massive gravity in unitary gauge
admits BH solutions  precisely of this  type.

Interestingly,  the dS-Schwarzschild solution found in  \cite {Koyama,KoyamaSA}
do happen to satisfy our conservative criterion of non-singularity.
However, the solution that we present  here is not among the ones
of \cite {Koyama,KoyamaSA}.

One more point worth emphasizing is that the BH 
solutions of \cite{GruzinovMirbabayi} do exhibit the ``helicity-0 hair'' 
(e.g., produce an extra scalar force), while the ones found in \cite {Koyama,KoyamaSA},  
and in the present work do not. The status of the ``no-hair'' theorems in GR with 
the galileon field (which should capture some properties of the  helicity-0 of 
massive gravity)  will be discussed in Ref. \cite {Lam}.

A limitation of our work is that we only manage to find these exact analytic solutions
for a   specific choice of the two  free parameters of massive gravity.
Such a choice is  peculiar since on the obtained  background, as we will show,
the kinetic terms for both the vector and scalar fluctuations vanish in the decoupling limit.

This fact would imply  infinitely strong interactions for these modes
(unless these modes happen to be nondynamical
to all orders,  e.g., due to the specific choice of the coefficients of the theory).
Because of this issue,  we would like  to regard the solutions  obtained here as just
examples demonstrating how non-singular solutions should emerge. We also hope that
our solutions are representative of a broader class of solutions which
may have better behaved fluctuations.

In this regard, there  seems to be a few directions in which the studies of  massive gravity
BH's can be extended. First, one could look at the metric  in the
unitary gauge which  would be  some generalization
of the    Kruskal-Szekeres form.  Second, one can extend the massive theory
of \cite {resum} by adding  more degrees of freedom
to the existing 5 helicity states of massive graviton.  In fact,
two consistent extensions have already been discussed so far: (I) adding one real
scalar field that makes the graviton
mass  dynamical \cite {nonFRW};   (II)  adding  one massless tensor field with two degrees
of freedom \cite {HRbi} that makes the internal space metric of the {\stu}  field dynamical (bigravity).
In the latter  case  cosmological solutions  were found
recently in \cite {VolkovII} and \cite {LuigiII},  while  BH's
were studied  in \cite {Pilo}.

\vspace{0.1in}

The  work is organized as follows. Section 2 gives a  brief review of the
theory of massive gravity \cite {resum}. In section 3  we find an
exact Schwarzschild-de Sitter solution,   and in section 4 an
exact Reissner-Nordstr\"om-de Sitter,  solution  which have nonsingular $I^{ab}$.
These solution are  similar to those  discovered by Koyama, Niz and Tasinato, and
by Th.~Nieuwenhuizen, but differ in detail: our dS solution has no ghost
even though the vector field is present (compare to  \cite {KoyamaSA}).
Moreover,  on the obtained solution the
singularities in the invariant $I^{ab}$ are absent (compare to  \cite {theo}).
In the Appendix A  we give another  exact Schwarzschild solution that
asymptotes to  a conformaly rescaled Minkowski space, and briefly mention its peculiarities.
In the Appendix B we discuss fluctuations  on the selfaccelerated solution of section 3.

\section{The Theory}

A massive graviton is described by the Lagrangian density of \cite {resum} specified below
\beq
\mathcal{L}=\frac{\mpl^2}{2}\sqrt{-g}\left(R + m^2 \mathcal U(g,\phi^a)\right),
\eeq
where $\mathcal{U}$ is the potential for the graviton that depends  on two  free
parameters $\alpha_{3,4}$
\beq
\mathcal U(g,\phi^a)=\(\mathcal{U}_2+\alpha_3\ \mathcal{U}_3+\alpha_4\ \mathcal{U}_4 \),
\label{eq:fullU}
\eeq
and the individual terms in the potential are defined as follows:
\beq
\label{po}
\mathcal{U}_2&=&[\mathcal{K}]^2-[\mathcal{K}^2]\,,\\
\mathcal{U}_3&=&[\mathcal{K}]^3-3 [\mathcal{K}][\mathcal{K}^2]+2[\mathcal{K}^3]\,,\\
\mathcal{U}_4&=&[\mathcal{K}]^4-6[\mathcal{K}^2][\mathcal{K}]^2+8[\mathcal{K}^3]
[\mathcal{K}]+3[\mathcal{K}^2]^2-6[\mathcal{K}^4]\,,
\label{pot}
\eeq
where $\mathcal{K}^\mu_\nu(g,\phi^a)=\delta^\mu_\nu-\sqrt{g^{\mu\alpha}\partial_\alpha \phi^a
\partial_\nu \phi^b \eta_{ab}}$;  rectangular brackets denote
traces, $[\mathcal{K}]\equiv {\rm Tr} (\mathcal{K})= \mathcal{K}^\mu_\mu$. The above potential is unique --
no further polynomial terms can be added to the action
without introducing the BD ghost.

The tensor $H_{\mu\nu}$ represents the covariantized metric perturbation, as discussed in the
introduction, which reduces  to the $h\mn$ in unitary gauge.
While in a gauge unfixed theory we have
\beq
H\mn=g\mn-\partial_\mu \phi^a \partial_\nu \phi^b \eta_{ab}.
\eeq
Moreover, $\mathcal{U}$ is constructed in such a way that the theory admits the Minkowski background
\beq
g_{\mu\nu}=\eta_{\mu\nu},\qquad \phi^a=x^\mu\delta_\mu^a.
\eeq
Hence, it is natural to split $\phi$'s as the background plus the `pion' contribution
$\phi^a=x^a-\pi^a,$ and as it
was already mentioned in the introduction,  the unitary gauge is defined by the condition $\pi^a=0$.
In the non-unitary gauge, on the other hand, it proves to be useful to adopt  the following decomposition
\beq
\pi^a=\frac{mA^a+\partial^a\pi}{\Lambda^3},
\label{decomp}
\eeq
where $A^\mu$ describes in the decoupling limit the helicity $\pm 1$, while $\pi$ is the longitudinal
mode of the graviton (in the  decoupling limit \cite{nima},
$\mpl\rightarrow \infty$ and $m\rightarrow 0$,  while $\Lambda^3\equiv\mpl m^2$ is held fixed).
This limit captures  the approximation in which the energy scale  is  much greater than
the graviton mass scale,  $E\gg m$.

For convenience, in what follows, we define the coefficients $\alpha$ and $\beta$ which are related
to those of \eqref{eq:fullU} by $\alpha_3=-(-\alpha+1)/3$ and $\alpha_4=-\beta/2+(-\alpha+1)/12$.
For generic values of the parameters $\alpha$ and $\beta$ the theory exhibits 
the Vainshtein mechanism, as show in the decoupling limit
\cite {general},  and beyond \cite {Koyama,DatoGiga}.  As was emphasized in \cite {general},
for one special choice, $\alpha =\beta=0$,  the nonlinear interactions vanish
in the decoupling limit with fixed $\Lambda$,  leaving the theory weakly coupled 
(i.e., no Vainshtein mechanism) in this limit.
For this particular choice of the coefficients the action of massive gravity with the potential 
\eqref{po}-\eqref{pot} (which can be rewritten in terms of just $[K]$ and tuned to it cosmological constant 
\cite {Rachel0}, referred as a minimal model  in Ref. \cite {Rachel0}),  was shown not 
to exhibit  the  Vainshtein mechanism also away from  the decoupling limit \cite {Koyama}.

\section{A Black Hole on de Sitter}

In this section we present the Schwarzschild-de Sitter solution in  the theory of
massive gravity described above. The obtained solution is free
of singularities (except from the conventional one appearing in GR).

For convenience we choose unitary gauge for the metric. In this gauge
the symmetric tensor $g_{\mu\nu}$ is  an observable  describing all the five degrees of
freedom of a massive graviton. The equations of motion in empty space read as follows
\beq
G_{\mu\nu}+m^2 X_{\mu\nu}=0\,,
\label{ein}
\eeq
where $X_{\mu\nu}$ is the effective energy-momentum tensor due to the graviton mass,
\beq
&&X\mn=-\frac12\Big[
\mathcal{K}g\mn-\K\mn+\alpha\(\K^2\mn-\K \K\mn+\frac 12 g\mn \([\K]^2-[\K^2]\)\)\\
&&+6 \beta\(\K\mn^3-\K \K\mn^2+\frac 12 \K\mn \([\K]^2-[\K^2]\)-\frac 16 g\mn \([\K]^3-3 [\K][\K^2]+2[\K^3]\)\)
\Big]\notag\,.
\eeq
Using the Bianchi  identities, from \eqref{ein} we obtain
the following constraint on the metric
\beq
m^2 \nabla^{\mu}X_{\mu\nu}=0,
\label{const}
\eeq
where  $\nabla ^\mu$ denotes the covariant derivative.

In order to obtain  the expression for $X\mn$  we make use of the fact that the
Lagrangian is written as the trace of the polynomial of the matrix $\mathcal{K}^\mu_\nu$.
Thus, following the method by Koyama, Niz and Tasinato \cite {Koyama}, we choose the basis which diagonalizes the expression appearing
under the square root in the  definition of $\mathcal{K}^\mu_\nu$ [One  should bear in mind that this is
not a coordinate transformation, but rather a trick to simplify the procedure of getting the
equations of motion].  As a result, the potential becomes a function of the components of the
inverse metric, rather than the combination of square roots of matrices. Having done this,
one is free to vary the action with respect to the inverse metric components to obtain explicit
expression for \eqref{ein}. Since these  expressions are quite cumbersome  we will not
give them here.

Below, we  concentrate on one particular  family  of the ghost-free theory of
massive gravity in which there is the following relations between the two free coefficients:
\beq
\beta=-\frac{\alpha^2}{6}\,.
\label{fam1}
\eeq
That this choice of the coefficients is special was first  shown by Th. Nieuwenhuizen \cite {theo} (see also
\cite{GruzinovMirbabayi}).  In particular, it was  shown in  \cite {theo} that for this choice the equation
\eqref{const} is automatically satisfied  for a certain diagonal (in spherical coordinates)
and time-independent metrics. It is interesting, however, that the above property
persists for a more general  class of non-diagonal spherically-symmetric metrics written as follows:
\beq
ds^2=-A(r)dt^2+2B(r)dtdr+C(r)dr^2+w^2r^2d\Omega^2,
\label{anz}
\eeq
where  $w$ is  a constant, while $A(r)$, $B(r)$ and $C(r)$ are arbitrary functions.

In subsection 3.1 we find an exact de Sitter solution to \eqref{ein},
and in subsection 3.2  we find an exact  BH solution on  the obtained
dS background. Note that the dS background is entirely due to the graviton mass.

\subsection{The de Sitter Solution}

We note that we would find an exact dS solution if we
required that
\beq
m^2 X\mn=\lambda g\mn,
\label{lambdacc}
\eeq
where  $\lambda$ is some constant. The solution of the equations \eqref{ein}
that also satisfies (\ref {lambdacc}) with a positive but otherwise arbitrary
$\alpha$ is given by
\beq
ds^2=-\kappa ^2 dt^2+\left(\frac{\alpha}{\alpha+1}dr\pm\kappa{\sqrt{\frac{2}{3\alpha}}
\frac{\alpha}{(\alpha+1)}mr} dt\right)^2+\frac{\alpha^2}{(\alpha+1)^2}r^2d\Omega^2.
\label{pain}
\eeq
Here, $\kappa$ is a positive integration constant. It is straightforward to check that
for  \eqref{pain}  we have $X_{\mu\nu}=(2/\alpha) ~g_{\mu\nu}$, leading to the expression for the Ricci scalar
\beq
R=\frac{8}{\alpha}m^2,
\eeq
as expected. Hence, this is a dS space with curvature scale
set  by the graviton mass and one free parameter $\alpha$. One could imagine that
$m\sim (0.1- 1) H_0$, and $\alpha \sim (0.01 -1)$, in which case
the obtained dS solution (if stable) could describe dark energy.

Up  to a rescaling of the coordinates,
the expression  \eqref{pain} looks exactly like the de Sitter solution  of GR
written in the Gullstrand--Painlev\'e frame. Either $\pm$ solution  covers half of dS space.
One can rotate the obtained
solution to the static coordinate system at the expense of
nonzero {\stu} fields. This will be done in the next subsection.
In either form, the solution has no additional singularities.

\subsection{Schwarzschild-de Sitter Background}

Having the solution of the previous subsection worked out, it is
straightforward to show that the system of equations \eqref{ein}
admits the following exact solution
\beq
ds^2=-\kappa ^2 dt^2+\left({\tilde \alpha} dr\pm\kappa\sqrt{\frac{r_g}{{\tilde \alpha} r}+
\frac{2{\tilde \alpha}^2 }{3\alpha} m^2r^2} dt\right)^2+ {\tilde \alpha}^2 r^2d\Omega^2\,,
\label{pain1}
\eeq
where ${\tilde \alpha } \equiv \alpha/(\alpha+1)$, and as before,
$\kappa$ is an integration constant.   In order to bring this solution to
a more familiar form  let us perform the following rescaling
\beq
r&\rightarrow& \frac{\alpha+1}{\alpha} r,\nonumber\\
dt&\rightarrow& \frac{1}{\kappa}dt.
\label{trans}
\eeq
The resulting metric reads
\beq
ds^2=-dt^2+\left(dr\pm\sqrt{\frac{r_g}{r}+\frac{2}{3\alpha}m^2r^2} dt\right)^2+r^2d\Omega^2.
\label{pain2}
\eeq
This is the Schwarzschild-de Sitter solution  in the GP coordinates.

However, the above rescaling  takes us away from the unitary gauge
\beq
&&\phi^0=t\rightarrow t-\left(1- \frac{1}{\kappa} \right) t,\\
&&\phi^r=r\rightarrow r+\frac{1}{\alpha} r.
\eeq
In terms of the  `pions',  $\pi^\mu\equiv x^\mu-\phi^\mu$, which can be decomposed as
$\pi^\mu=(mA^\mu+\partial^\mu\pi)/\Lambda^3$, we have
\beq
\pi&=&\frac{\Lambda^3}{2}\left[-\left( 1-\frac{1}{\kappa} \right)t^2-\frac{1}{\alpha}r^2\right],\\
A^\mu&=&0.
\label{asig}
\eeq
The fields in (\ref {asig}) correspond to the canonically normalized fields carrying
the helicity eigenstates in the decoupling limit.

\vspace{0.1in}

Let us now rewrite our solution into a more familiar coordinate system.
The metric can be transformed  to a  static slicing by means of the following
coordinate transformation
\beq
dt\rightarrow dt+f'(r)dr,
\label{trans1}
\eeq
with $f'(r)\equiv -g_{01}/g_{00}$ given by
\beq
f'(r)=\pm \frac{\sqrt{\frac{r_g}{r}+\frac{2}{3\alpha}m^2r^2}}{1-\frac{r_g}{r}-\frac{2}{3\alpha}m^2r^2}.
\eeq
The resulting expression for the  metric reads as follows:
\beq
ds^2=-\left(1-\frac{r_g}{r}-\frac{2}{3\alpha}m^2r^2\right)dt^2+\frac{dr^2}
{1-\frac{r_g}{r}-\frac{2}{3\alpha}m^2r^2}+r^2d\Omega^2.
\label{static}
\eeq
This is nothing but the metric of the Schwarzschild-de Sitter
solution of GR  in the static coordinates. However, this metric  should be accompanied
by a nontrivial  backgrounds for the {\stu} fields. Indeed,  it is evident that \eqref{trans1}
gives rise to  the shift $\delta \phi^0=f(r)$.  In turn, this gives rise to
a background for the `vector mode'
\beq
A^0&=&-\frac{\Lambda^3}{\kappa m}f(r),\\
A^i&=&0.
\label{asig0}
\eeq
This particular field assignment has been chosen according to scaling in
the decoupling limit. Namely, $f(r)$ vanishes in the decoupling limit linearly in $m$
hence it was ascribed to the ``vector mode''.

\vspace{0.1in}

We would like to make two important comments  in the remainder of this section.
The first one concerns the integration constant $\kappa$.
Although, all the invariants of GR  are independent of $\kappa$,
the new invariant  that is characteristic of  massive gravity
\beq
I^{ab}\equiv g^{\mu\nu}\partial_\mu \phi^a \partial_\nu \phi^b,
\label{newinv}
\eeq
does  depend on this integration constant; in the unitary gauge  $I^{ab}$ is just
the inverse of the GP metric \eqref{pain}  which reads as follows
\beq
\footnotesize{\left(
 \begin{array}{ccc} \nonumber
-\frac{1}{\kappa^2} & \pm \frac{1}{\kappa} \frac{\alpha+1}{\alpha}\sqrt{\frac{\alpha+1}{\alpha}\frac{r_g}{r}+\frac{2}{3\alpha}\frac{\alpha^2}{(\alpha+1)^2}m^2r^2} & 0 \\
\pm \frac{1}{\kappa} \frac{\alpha+1}{\alpha}\sqrt{\frac{\alpha+1}{\alpha}\frac{r_g}{r}+\frac{2}{3\alpha}\frac{\alpha^2}{(\alpha+1)^2}m^2r^2} & \left( \frac{\alpha+1}{\alpha} \right)^2\left( 1-\frac{\alpha+1}{\alpha}\frac{r_g}{r}-\frac{2}{3\alpha}\frac{\alpha^2}{(\alpha+1)^2}m^2r^2 \right) & 0 \\
0 & 0 & \left( \frac{\alpha+1}{\alpha} \right)^2 \Omega^{-1}_{2\times 2}
\end{array}
\right)}.
\eeq
Thus, the backgrounds with different values of $\kappa$ correspond to distinct super-selection sectors
labeled by the values of $I^{ab}$.

The second comment concerns the issue of small fluctuations on top of this solution.
One may worry that the scalar perturbations on this  background  are infinitely strongly
coupled in the light of the results of \cite{dato}. In the latter work  it was found that,
for the parameters chosen as in \eqref{fam1}, the de Sitter background has
infinitely strongly coupled fluctuations in the decoupling limit.    However,  we should point out that
the self-accelerated  background discussed in this section is different from that
studied in \cite{dato}. This distinction is manifest in \eqref{asig0} by the presence of the
background for $A_0$, which vanishes in the case of \cite{dato}.

Still,  one could argue that it is unnecessary to perform the transformation of
variables \eqref{trans1} responsible for this difference, and limit oneself
to the rescaling of the  coordinates
\beq
r \rightarrow  \frac{\alpha+1}{\alpha} r\,,~~~t \rightarrow  \frac{1}{\kappa}t\,,
\label{hyp}
\eeq
which clearly does not give  rise to the vector background.
As a result,  the `pion' configuration will become similar to that  of \cite{dato}, while  the metric itself
will be quite different, namely\footnote{For simplicity we set $r_g=0$ and drop the numerical factors.}
\beq
ds^2=-(1-m^2r^2)dt^2+2mrdtdr+dr^2+r^2d\Omega^2.
\label{met}
\eeq
Now, if we were to take this metric as the one in which  the decoupling limit should be taken, then
we would  find that the gauge freedom that is left in this limit
\beq
g\mn \rightarrow g\mn+\partial _{( \mu} \xi _{\nu )},
\eeq
would  not be enough for bringing \eqref{met} to the form of de Sitter space in either
conformal or static slicing. Furthermore, the canonically normalized \eqref{met} diverges in the decoupling limit, in such a way that this divergence can be isolated only in the vector mode. If so, then
no conclusion can be drawn about the perturbations around our background based on the
results of  \cite{dato}.  This,  on the other hand,  does not necessarily imply that the
fluctuations are fine.  As we show in the Appendix B the vector and scalar
fluctuations  may be infinitely strongly coupled.

\subsection{From Gullstrand-Painlev\'e to Kruskal-Szekeres}

In this section we ask the question whether the BH solution of the GP form could be analytically continued to cover the other half of the space-time as well. This can be done by going to the Kruskal-Szekeres (KS) coordinates and analyzing the {\stu} fields.

Let us start addressing this point by considering the following background
\beq
ds^2=-dt_{GP}^2+\left(dr+\sqrt{\frac{r_g}{r}}dt_{GP}\right)^2+r^2 d\Omega^2, \qquad \phi^a=x^a.
\label{metric}
\eeq
First we rewrite the metric in static slicing by performing the following change of the time variable
\beq
\phi^0=t_{GP}=t+2r_g \sqrt{\frac{r}{r_g}}+r_g \text{ ln}\left( \frac{\sqrt{\frac{r}{r_g}}-1}{\sqrt{\frac{r}{r_g}}+1} \right).
\eeq
As a result the metric takes on the Schwarzschild form.
In order to go to KS coordinates we use reparametrizations identical to the one used in GR
\beq
&&\left(\frac{r}{r_g}-1\right)e^{r/r_g}=X^2-T^2, \\
&&t=r_g\text{ ln}\left( \frac{X+T}{X-T} \right).
\eeq
For the analysis of the $\phi$'s in KS coordinates we make the `near the horizon' approximation $r/r_g\rightarrow 1$, since this is the the region of our interest. In this limit the above coordinate transformation simplifies to (near the horizons $T=\pm X$, with signs corresponding to the black- and white-hole respectively)
\beq
&&\left(\frac{r}{r_g}-1\right)=\frac{1}{e}(X^2-T^2), \\
&&t=r_g\text{ ln}\left( \frac{X+T}{X-T} \right).
\eeq
As a result the $\phi$'s take the following form (using the fact that $T^2-X^2$ is small)
\beq
&&\phi^0=2r_g\text{ ln}(X+T)+r_g (\text{ln}(1/4)+1),\\
&&\phi^r=r_g\left(1+\frac{1}{e}(X^2-T^2)\right).
\eeq
Notice that $\phi^0$ is singular on the horizon of the white hole (while being regular on the black hole horizon).
The metric in these coordinates is given by
\beq
ds^2=4r_g^3\frac{e^{-r/r_g}}{r}(-dT^2+dX^2)+r^2d\Omega^2
\eeq
The invariant $I^{ab}=g^{\mu\nu}\partial_\mu \phi^a \partial_\nu \phi^b$ on the above background is singular at $X=-T$, corresponding to the horizon of the white hole.

If one takes the original GP metric \eqref{metric} to describe the white hole instead of the black hole (this is achieved by flipping the relative sign of the expressions in parentheses) then after analytical continuation to KS coordinates the singularity will appear on the black hole horizon rather than on the one of the white hole. The generalization of this arguments for the case of the dS is straightforward.

\section{Reissner-Nordstr\"om solution on de Sitter}

The ghost-free theory of massive gravity with $\beta=-\alpha^2/6$, upon its coupling to the Maxwell's theory of electromagnetism, possesses the following Reissner-Nordstr\"om solution on dS space
\beq
ds^2=-dt^2+\left(\tilde{\alpha}dr\pm\sqrt{\frac{r_g}{\tilde{\alpha}r}+\frac{2\tilde{\alpha}^2}{3\alpha} m^2r^2-\frac{\tilde{Q}^2}{\tilde{\alpha}^4 r^2}} dt\right)^2+\tilde{\alpha}^2 r^2d\Omega^2,
\label{GPcharged}
\eeq
with $\tilde{\alpha} \equiv \alpha/(\alpha+1)$ and the electromagnetic field given by
\beq
E=\frac{\tilde{Q}}{r^2} \qquad \text{and} \qquad B=0.
\eeq
In order to normalize the radial coordinate appropriately and to rewrite the solution in the static slicing, one needs to perform the rescaling
\beq
r\rightarrow \frac{r}{\tilde{\alpha}},
\eeq
supplemented with the following transformation of time
\beq
dt\rightarrow dt+f'(r)dr, \quad \text{with} \quad f'(r)\equiv -\frac{g_{01}}{g_{00}}=\pm \frac{\sqrt{\frac{r_g}{r}+\frac{2}{3\alpha} m^2r^2-\frac{\tilde{Q}^2}{\tilde{\alpha}^2 r^2}}}{1-\frac{r_g}{r}-\frac{2}{3\alpha} m^2r^2+\frac{\tilde{Q}^2}{\tilde{\alpha}^2 r^2}}.
\eeq
As a result the metric takes the familiar form
\beq
ds^2=-(1-\frac{r_g}{r}-\frac{2}{3\alpha} m^2r^2+\frac{\tilde{Q}^2}{\tilde{\alpha}^2 r^2})dt^2+\frac{dr^2}{1-\frac{r_g}{r}-\frac{2}{3\alpha} m^2r^2+\frac{\tilde{Q}^2}{\tilde{\alpha}^2 r^2}}+r^2d\Omega^2,
\eeq
while the {\stu} fields become
\beq
\phi^0&=&t+f(r), \\
\phi^r&=&r+\frac{1}{\alpha} r.
\eeq
In this reference frame the electromagnetic field is
\beq
E=\frac{\tilde{Q}}{\tilde{\alpha} r^2} \qquad \text{and} \qquad B=0.
\eeq
And, for obvious reasons the actual charge should be defined by $Q\equiv \tilde{Q}/\tilde{\alpha}$.

\vskip 10pt

\subsection*{Acknowledgments}

We'd like to thank Gia Dvali, Lam Hui, Mehrdad Mirbabayi, Alberto Nicolis,  
and David Pirtskhalava for useful comments. The work of LB and GC are  supported by 
the NYU James Arthur and MacCraken Fellowships, respectively.  CdR is supported by the 
Swiss National Science Foundation.  GG is supported by NSF grant PHY-0758032. AJT would like to thank the
Universit\'e de  Gen\`eve for hospitality whilst this work was being completed.


\renewcommand{\theequation}{\Roman{equation}}
\setcounter{equation}{0}

\appendix

\section{Schwarzschild-like  Solution}

In this appendix  we concentrate on a different choice of the parameters
\beq
\beta=-\frac{\alpha^2}{8},
\label{fam2}
\eeq
for which some exact solutions can also be obtained. In particular, we show
that there exists an exact non-singular and asymptotically flat BH solution. We choose the unitary gauge and
find that $X_{tt}$ and $X_{tr}$ from eq.
\eqref{ein} vanish  identically on the ansatz \eqref{anz}, for $w=\alpha/(\alpha+2)$. Suggesting, that there exist fluctuations which are infinitely strongly coupled.

Then, the solution to the full set of the equations of motion takes the following form
\beq
\begin{split}
ds^2 & =-\frac{(\alpha+2)^3\alpha^2}{(\alpha+2)^5+\alpha^5\delta} \left( 1-\frac{r_g(\alpha+2)}{r\alpha}\right)dt^2
\\ & \pm \frac{2\alpha(\alpha+2)}{(\alpha+2)^5+\alpha^5\delta}\sqrt{\frac{r_g^2(\alpha+2)^6}{r^2}+\frac{r_g\alpha^6\delta}{r}}dtdr\\&+ \frac{\alpha^2}{(\alpha+2)^2} \left(1+\frac{r_g(\alpha+2)^6}{\alpha r((\alpha+2)^5+\alpha^5\delta)} \right)dr^2+\frac{\alpha^2 r^2}{(\alpha+2)^2}d\Omega^2\,,
\label{schp}
\end{split}
\eeq
where $\delta$ and $r_g$ are positive integration constants. The transformation to the Schwarzschild
coordinates is carried out in a  way similar to the previous section
\beq
r&\rightarrow& \frac{\alpha+2}{\alpha} r,\nonumber\\
dt&\rightarrow& \zeta \left(dt+f'(r)dr\right), \quad \text{with}\quad \zeta^2\equiv\frac{(\alpha+2)^2}{\alpha^2}+\frac{\alpha^3}{(\alpha+2)^3}\delta.
\eeq
As a result the metric takes the conventional  form
\beq
ds^2=-\left( 1-\frac{r_g}{r}\right)dt^2+\frac{dr^2}{1-\frac{r_g}{r}}+r^2 d\Omega^2,
\label{bh}
\eeq
however, this should be accompanied by the  `pion' configuration
\beq
\pi&=&\frac{\Lambda^3}{2}\left[-\left( 1-\zeta \right)t^2-\frac{2}{\alpha}r^2\right],\\
A^0&=&-\zeta \frac{\Lambda^3}{m}f(r),\\
A^i&=&0.
\label{asig1}
\eeq
It is interesting  that there exists a choice of parameters for which one gets the background metric
identical to that of GR, supplemented with the `pion' fields listed above.
All the invariants are regular on this solution (away from the  singularity in the center).
It should be pointed out that \eqref{schp} is the  only background among those given
in \cite{Koyama} with vanishing cosmological constant. This solution, however, has infinitely strongly coupled
fluctuations; this and related issues will be discussed in a forthcoming paper \cite {prep}.

\section{Fluctuations}

In this section we study the fluctuations on the backgrounds of section 3.
The analysis is done in the decoupling limit \cite {nima}
\beq
m\rightarrow 0, \qquad \mpl \rightarrow \infty, \quad \text{with} \quad \Lambda^3 \equiv\mpl m^2 - \text{fixed}.
\eeq
In this limit the Lagrangian  can be decomposed into tow pieces.
The first describes the dynamics of the helicity-$0,\pm 2$ modes and their interactions with each other.
While the second one accounts for the helicity-$\pm1$ modes and their nonlinear
couplings to the helicity-0 degree of freedom.

The scalar-tensor Lagrangian, with the condition $\beta = -\alpha^2/6$,   is given by \cite {general}
\beq
\mathcal{L}_{ST}=-\frac{1}{2} h^{\mu \nu}\mathcal{E}^{\alpha \beta}_{\mu \nu}h_{\alpha \beta}+
h^{\mu \nu} \left( X^{(1)}\mn-\frac{\alpha}{\Lambda^3}X^{(2)}\mn-\frac{\alpha^2}{6\Lambda^3}X^{(3)}\mn \right).
\label{ST}
\eeq
Here, the first term represents the linearized Einstein-Hilbert Lagrangian,
while $X$'s are defined in terms  of the longitudinal mode  as follows
\beq
X^{(1)}\mn&=&-\frac{1}{2} \epsilon_{\mu \alpha \rho \sigma}\epsilon_{\nu \beta \rho \sigma}\Pi_{\alpha \beta},\nonumber\\
X^{(2)}\mn&=&\frac{1}{2} \epsilon_{\mu \alpha \rho \gamma}\epsilon_{\nu \beta \sigma \gamma}\Pi_{\alpha \beta}\Pi_{\rho \sigma},\nonumber\\
X^{(3)}\mn&=&\epsilon_{\mu \alpha \rho \gamma}\epsilon_{\nu \beta \sigma \delta}\Pi_{\alpha \beta}\Pi_{\rho \sigma}\Pi_{\gamma \delta},\nonumber
\eeq
with $\Pi_{\mu \nu}\equiv \partial _\mu \partial _\nu \pi$ and all the repeated indices
contructed by the flat space metric.

The scalar-vector Lagrangian, on the other hand, contains an infinite number of terms
and schematically is given by
\beq
\mathcal{L}_{SV}=-\frac{1}{4} F\mn^2+\displaystyle\sum\limits_{n=1}^\infty \partial
A \partial A \left(\frac{\partial \partial \pi}{\Lambda^3}\right)^n,
\eeq
where $F\mn$ denotes the field strength of the  helicity-1 mode, $A_\mu$, and the whole Lagrangian is invariant under the $U(1)$ gauge transformation $\delta A_\mu=\partial_\mu \alpha$. Remarkably,
this expression has been recently resummed for the spherically symmetric ansatz \cite{KoyamaSA},
making the analysis of the vector fluctuations possible.

After expanding $\mathcal{L}_{ST}$,  and the resummed version of $\mathcal{L}_{SV}$
(eq. (C.5) of Ref. \cite{KoyamaSA} which is too lengthy to be reproduced here)
to the second order in perturbations around the backgrounds
of section 3, we find that the  kinetic terms for both  helicity-$0$ and helicity-$\pm1$ fields
vanish identically, on the dS as well as the Schwarzschild-dS space.
This does imply that these modes are infinitely strongly coupled in
this limit (unless of course they are rendered nondynamical  to all orders by some symmetry
or constraint). Whether this problem can be remedied by going beyond the decoupling
limit, and/or by invoking kinetic terms due
to quantum loops (which will be generated as long as they're not
prohibited by symmetries), remains to be seen.


\begin{thebibliography}{99}

\bibitem{FP}
  M.~Fierz and W.~Pauli,
  Proc.\ Roy.\ Soc.\ Lond.\  A {\bf 173}, 211 (1939).

  \bibitem{Nieu}  P.~van Nieuwenhuizen,
  Nucl.\ Phys.\  B {\bf 60} (1973) 478.

\bibitem{vDVZ}
  H.~van Dam and M.~J.~G.~Veltman,
  Nucl.\ Phys.\  B {\bf 22}, 397 (1970);

\bibitem{Arkady}
  A.~I.~Vainshtein,
  Phys.\ Lett.\  B {\bf 39}, 393 (1972).

  \bibitem{dvali}
  C.~Deffayet, G.~R.~Dvali, G.~Gabadadze and A.~I.~Vainshtein,
  Phys.\ Rev.\  D {\bf 65}, 044026 (2002)
  [arXiv:hep-th/0106001].

\bibitem{BD}
  D.~G.~Boulware and S.~Deser,
  Phys.\ Rev.\  D {\bf 6}, 3368 (1972).

\bibitem{nima}
  N.~Arkani-Hamed, H.~Georgi, M.~D.~Schwartz,
  Annals Phys.\  {\bf 305}, 96-118 (2003).
  [hep-th/0210184].


  \bibitem{Creminelli}
  P.~Creminelli, A.~Nicolis, M.~Papucci and E.~Trincherini,
  JHEP {\bf 0509}, 003 (2005).

\bibitem{DeffayetRombouts}
  C.~Deffayet and J.~W.~Rombouts,
  Phys.\ Rev.\  D {\bf 72}, 044003 (2005)
  [arXiv:gr-qc/0505134].

\bibitem{Andrei}
  G.~Gabadadze and A.~Gruzinov,
  Phys.\ Rev.\  D {\bf 72}, 124007 (2005)
  [arXiv:hep-th/0312074].



\bibitem{general}
  C.~de Rham, G.~Gabadadze,
  Phys.\ Rev.\  {\bf D82}, 044020 (2010).
  [arXiv:1007.0443 [hep-th]].
  

\bibitem{resum}
  C.~de Rham, G.~Gabadadze, A.~J.~Tolley,
   Phys.\ Rev.\ Lett.  {\bf 106}, 231101 (2010).
[arXiv:1011.1232 [hep-th]].



\bibitem{rachel1}
 S.~F.~Hassan, R.~A.~Rosen,
    [arXiv:1106.3344 [hep-th]].



\bibitem{Lham}
J.~Kluson,
  arXiv:1109.3052 [hep-th].

\bibitem{Rachel_confirm} 
S.~F.~Hassan, R.~A.~Rosen,
  [arXiv:1111.2070 [hep-th]].



  \bibitem{Stu} C.~de Rham, G.~Gabadadze and A.~Tolley,
  arXiv:1107.3820 [hep-th].


\bibitem{Helic} C.~de Rham, G.~Gabadadze and A.~J.~Tolley,
  arXiv:1108.4521 [hep-th].


\bibitem{Siegel} W.~Siegel,
  Phys.\ Rev.\  {\bf D49}, 4144-4153 (1994).
  [hep-th/9312117].



\bibitem{sergey}
S.~L.~Dubovsky,
  JHEP {\bf 0410}, 076 (2004)
  [arXiv:hep-th/0409124].


\bibitem{CedricJ}
C.~Deffayet and T.~Jacobson,
  arXiv:1107.4978 [gr-qc].


\bibitem{Koyama}
K.~Koyama, G.~Niz, G.~Tasinato,
  Phys.\ Rev.\ Lett.\  {\bf 107}, 131101 (2011).
  [arXiv:1103.4708 [hep-th]],

K.~Koyama, G.~Niz, G.~Tasinato,
  Phys.\ Rev.\  {\bf D84}, 064033 (2011).
  [arXiv:1104.2143 [hep-th]].

\bibitem{theo}
  T.~M.~Nieuwenhuizen,
  Phys.\ Rev.\  D {\bf 84}, 024038 (2011)
  [arXiv:1103.5912 [gr-qc]].

\bibitem{GruzinovMirbabayi}
A.~Gruzinov and M.~Mirbabayi,
  arXiv:1106.2551 [hep-th].

\bibitem{dato}
  C.~de Rham, G.~Gabadadze, L.~Heisenberg, D.~Pirtskhalava,
  Phys.\ Rev.\  {\bf D83}, 103516 (2011).
  [arXiv:1010.1780 [hep-th]].


\bibitem{Rachel0} 
S.~F.~Hassan, R.~A.~Rosen,
  JHEP {\bf 1107}, 009 (2011).
  [arXiv:1103.6055 [hep-th]].


\bibitem{DatoGiga} G.~Chkareuli, D.~Pirtskhalava,
  [arXiv:1105.1783 [hep-th]].


\bibitem{Volkov}
 A.~H.~Chamseddine, M.~S.~Volkov,
  Phys.\ Lett.\  {\bf B704}, 652-654 (2011).
  [arXiv:1107.5504 [hep-th]].

\bibitem{nonFRW}
G.~D'Amico, C.~de Rham, S.~Dubovsky, G.~Gabadadze, D.~Pirtskhalava and A.~J.~Tolley,
  arXiv:1108.5231 [hep-th].

\bibitem{IPMU}
A.~E.~Gumrukcuoglu, C.~Lin and S.~Mukohyama,
  arXiv:1109.3845 [hep-th].


\bibitem{Mohseni}
M.~Mohseni, Phys.\ Rev.\ D {\bf 84}, 064026 (2011)
  [arXiv:1109.4713 [hep-th]].


\bibitem{VolkovII}
M.S. Volkov,
arXiv:1110.6153 [hep-th].



\bibitem{LuigiII}
D.~Comelli, M.~Crisostomi, F.~Nesti, L.~Pilo,
  [arXiv:1111.1983 [hep-th]].




\bibitem{KoyamaSA}
K.~Koyama, G.~Niz, G.~Tasinato,
  [arXiv:1110.2618 [hep-th]].

\bibitem{Lam} L. Hui, A. Nicolis, to appear.


\bibitem{HRbi} S.~F.~Hassan, R.~A.~Rosen,
  [arXiv:1109.3515 [hep-th]].


\bibitem{Pilo} D.~Comelli, M.~Crisostomi, F.~Nesti, L.~Pilo,
  [arXiv:1110.4967 [hep-th]].



\bibitem{Wyman:2011mp}
  M.~Wyman,
  Phys.\ Rev.\ Lett.\  {\bf 106}, 201102 (2011).
  [arXiv:1101.1295 [astro-ph.CO]].



\bibitem{prep}
In preparation.


\end{thebibliography}
\end {document}